\overfullrule=0pt
\input harvmac
\def\bar{\overline}
\def\a{{\alpha}}

\def\ah{{\widehat \a}}
\def\lh{{\widehat \lambda}}

\def\bd{{\dot \b}}
\def\l{{\lambda}}

\def\b{{\beta}}
\def\bh{{\widehat\beta}}
\def\dh{{\widehat\delta}}
\def\g{{\gamma}}
\def\gh{{\widehat\gamma}}

\def\d{{\delta}}
\def\e{{\epsilon}}
\def\s{{\sigma}}

\def\p{{\partial}}

\def\pb{{\overline\partial}}
\def\t{{\theta}}

\def\th{{\widehat\theta}}

\Title{\vbox{\baselineskip12pt
\hbox{IFT-P.011/2008 }}}
{{\vbox{\centerline{ Perturbative Super-Yang-Mills from the
 }
\smallskip
\centerline{Topological $AdS_5\times S^5$ Sigma Model}}} }
\bigskip\centerline{Nathan Berkovits\foot{e-mail: nberkovi@ift.unesp.br}}
\bigskip
\centerline{\it Instituto de F\'\i sica Te\'orica, S\~ao Paulo State University 
}
\centerline{\it Rua Pamplona 145, 01405-900, S\~ao Paulo, SP, Brasil}
\bigskip

\vskip .3in

A topological sigma model based on the pure spinor formalism
was recently proposed for the small radius limit
of the $AdS_5\times S^5$ superstring. Physical states in this model
can be constructed by connecting holes on the worldsheet with Wilson lines
of the worldsheet gauge field. The contribution of these states to
the topological amplitude is claimed to reproduce the usual Feynman diagram 
expansion of gauge-invariant super-Yang-Mills correlation functions.

\vskip .3in

\Date {June 2008}

\newsec{Introduction}

The pure spinor formalism can be used to covariantly
describe the superstring in
any consistent $d=10$ supergravity background \ref\pureone{N. Berkovits,
{\it Super-Poincar\'e Covariant Quantization of the Superstring},
JHEP 0004 (2000) 018, hep-th/0001035.}. When the supergravity
background is $AdS_5\times S^5$, the resulting worldsheet action has
manifest $PSU(2,2|4)$ symmetry and is constructed from the Metsaev-Tseytlin
left-invariant currents $g^{-1}dg$ where $g$ takes values in the coset
${{PSU(2,2|4)}\over{SO(4,1)\times SO(5)}}$ \ref\metsaev{R.R. Metsaev
and A.A. Tseytlin, {\it Type IIB Superstring Action in $AdS_5\times S^5$
Background}, Nucl. Phys. B533 (1998) 109, hep-th/9805028.}.
In the large radius limit where $r_{AdS}\to\infty$, this action can
be covariantly quantized \ref\adsquant{N. Berkovits and O. Chand\'{\i}a,
{\it Superstring Vertex Operators in an $AdS_5\times S^5$ Background},
Nucl. Phys. B596 (2001) 185, hep-th/0009168.}\ref\adsqt{N. Berkovits,
{\it Quantum Consistency of the Superstring in $AdS_5\times S^5$
Background}, JHEP 0503 (2005) 041, hep-th/0411170.}
and one can compute $PSU(2,2|4)$-covariant
correlation functions as an expansion in $1\over{r_{AdS}}$ \ref\mikh
{A. Mikhailov and S. Schafer-Nameki, {\it Perturbative Study of the
Transfer Matrix on the String Worldsheet in $AdS_5\times S^5$},
arXiv:0706.1525[hep-th].}. However,
to compare with computations in perturbative super-Yang-Mills, one
needs to be able to quantize the worldsheet action in the small radius
limit where $r_{AdS}\to 0$.

Recently, a proposal was made for how to quantize in the small radius limit
\ref\limone{N. Berkovits,
{\it New Limit of the $AdS_5\times S^5$ Sigma Model}, JHEP 0708 (2007) 011, 
hep-th/0703282.}\ref\limtwo{N. Berkovits and C. Vafa,
{\it Towards a Worldsheet Derivation of the Maldacena Conjecture},
JHEP 0803 (2008) 031, arXiv:0711.1799[hep-th].}.
After combining the 22 pure spinor ghosts $\l^\a$ and $\lh^\ah$
with the ten $AdS_5\times S^5$
spacetime variables into a 32-component unconstrained bosonic spinor,
the $AdS_5\times S^5$ worldsheet action was expressed as an $N=(2,2)$
worldsheet supersymmetric action based on the fermionic coset
${{PSU(2,2|4)}\over{SU(2,2)\times SU(4)}}$. This coset contains 32 
fermionic variables, and the 32-component
unconstrained bosonic spinor is the
worldsheet superpartner of these variables.

If the BRST charge is defined to be the scalar worldsheet supersymmetry
generator, this worldsheet supersymmetric action is a topological
A-model which can be quantized using standard topological methods.
However, in the large radius limit, it is important to note that
the BRST charge defined in the pure spinor formalism is {\it not} the
scalar worldsheet supersymmetry generator. So in the large radius limit,
the $AdS_5\times S^5$ worldsheet action is not a topological A-model,
which is expected since one has a continuum of physical states in
this supergravity limit.

Nevertheless, it was conjectured that in the small radius limit,
the BRST charge can be defined to be the scalar worldsheet supersymmetry
generator such that 
the worldsheet action for the $AdS_5\times S^5$ superstring
becomes a topological A-model when $r_{AdS}\to 0$.
Preliminary evidence for this conjecture came from an analogy with the
Gopakumar-Vafa duality relating $d=3$ Chern-Simons theory and the
resolved conifold \ref\gop{R. Gopakumar and C. Vafa,
{\it Topological Gravity as Large N Topological Gauge Theory}, 
Adv. Theor. Math. Phys. 2 (1998)
413, hep-th/9802016.}. This open-closed duality was proven
in \ref\ova{H. Ooguri and C. Vafa,
{\it Worldsheet Derivation of a Large N Duality}, Nucl. Phys. 
B641 (2002) 3, hep-th/0205297.}\ 
using a topological A-model
and has many similarities with super-Yang-Mills/$AdS_5\times S^5$
duality.
More recently, additional evidence for the conjecture was provided
by Bonelli and Safaai \ref\bonelli{
G. Bonelli and H. Safaai, {\it On Gauge/String Correspondence and
Mirror Symmetry}, arXiv:0804.2629[hep-th].}\
who argued that topological amplitudes involving
certain D-branes in the model compute correlation functions of circular
super-Yang-Mills Wilson lines. These D-branes break $PSU(2,2|4)$ to
$OSp(2,2|4)$ which are the symmetries preserved by the circular
Wilson lines.

If the conjecture is correct that this topological A-model describes
the small radius limit of $AdS_5\times S^5$, it should be possible
to compute correlation functions of arbitrary gauge-invariant
super-Yang-Mills operators using topological string methods. In
this paper, it will be argued that topological amplitudes in this model
indeed can compute arbitrary gauge-invariant
super-Yang-Mills correlation functions. The
topological amplitudes reproduce
the usual perturbative Feynman diagram method for computing these correlation
functions by replacing the propagators and vertices of
Feynman diagrams with a network of Wilson lines of a worldsheet gauge
field which connect holes on the closed string worldsheet.

The first step in computing these topological amplitudes is to note
that the BRST-invariant topological A-model of \limone\limtwo\ can be expressed
as the gauge-fixed version of a ${\cal G}/{\cal G}$ principal
chiral model where ${\cal G} = PSU(2,2|4)$.\foot{Based on analysis
using the RNS formalism,
a similar topological description of the zero radius limit of the $AdS_5\times
S^5$ superstring was discussed by Polyakov
at Strings 2002 \ref\polya{A. Polyakov, {\it Old and New Aspects of the Strings/Gauge
Correspondence}, Strings 2002 proceedings, 
http://www.damtp.cam.ac.uk/strings02/avt/polyakov/.}.}
This principal chiral model is defined by the worldsheet action
\eqn\actone{S = Tr\int d^2 z [r_{AdS}^2(g^{-1}\bar\nabla g)(g^{-1}\nabla g)
+ {1\over{e^2}} F^2 ]}
where $g$ takes values in $PSU(2,2|4)$, the covariant derivative on $g$
is gauged using a $PSU(2,2|4)$ worldsheet gauge field $(A,\bar A)$
whose field-strength
is $F$, and the infrared limit $e\to \infty$ is taken at the end of
the computation. 

If $r_{AdS}$ is large, one can freely set ${1\over{e^2}}=0$ and the model
becomes trivial by gauging away $g$ such that the action reduces
to $S=Tr\int d^2 z ~r^2_{AdS} A \bar A$. However, when $r_{AdS}$ is small,
there can be non-trivial fluctuations of the gauge field that survive in
the limit where $e\to\infty$. These fluctuations are of size
$(e~r_{AdS})^{-1}$ and can be described by closed string vertex operators
on the worldsheet which are connected to each other by a network of
Wilson lines.

For the configuration corresponding to $M$ gauge-invariant super-Yang-Mills
operators,
one will have $M$ vertex operators on the closed string worldsheet.
And if the $r^{th}$ gauge-invariant operator is
$Tr(\Phi_1 ...\Phi_{n_r})$ where $\Phi_1 ... \Phi_{n_r}$ are linearized
super-Yang-Mills fields, there will be $n_r$ Wilson lines emerging
from the $r^{th}$ hole which join with the Wilson lines emerging
from the other holes. 
This network of Wilson lines will represent a Feynman diagram
of perturbative super-Yang-Mills, 
and it will be required that Wilson lines do
not cross on the worldsheet so that the Feynman diagram can be
thickened as in the `t Hooft large $N$ expansion.
Furthermore, it will be claimed that the contribution of each network
to the topological amplitude coincides with the corresponding Feynman
rules including the factor of
$(\l_{YM}^2)^{2g-2}(\l_{'t Hooft})^{faces} = 
(\l_{string})^{2g-2}(r_{AdS}^4)^{faces}$
which is predicted by the Maldacena conjecture \ref\malda
{J. Maldacena, {\it
The Large N Limit of Superconformal Field Theories and Supergravity}, 
Adv. Theor. Math. Phys.
2 (1998) 231, hep-th/9711200.}.

Note that in the topological amplitude computation, there is no
integration over the locations of the closed string vertex operators. 
Unlike the proposal of \ref\gads{R. Gopakumar, {\it From Free Fields to
AdS}, Phys. Rev. D70 (2004) 025009, hep-th/0308184\semi
R. Gopakumar, {\it From Free Fields to
AdS: II}, Phys. Rev. D70 (2004) 025010, hep-th/0402063\semi
R. Gopakumar, {\it From Free Fields to
AdS: III}, Phys. Rev. D72 (2005) 066008, hep-th/0504229.}\
where the Schwinger parameters come
from integration over worldsheet moduli, integrals over loop momenta
in this description come from summing over the components in
the singleton representation of $PSU(2,2|4)$ which describe the propagating
states in the Feynman diagram. This is similar to
computations in twistor-string theory \ref\twist{E. Witten,
{\it Peturbative Gauge Theory as a String Theory in Twistor Space},
Comm. Math. Phys. 252 (2004) 189, hep-th/0312171.}
\ref\altern{N. Berkovits, {\it An Alternative String Theory
in Twistor Space for $N=4$ Super-Yang-Mills}, Phys. Rev. Lett.
93 (2004) 011601, hep-th/0402045.} where tree-level super-Yang-Mills
amplitudes are reproduced without any integration over worldsheet
moduli.

An interesting question is how these topological amplitude computations
are related to the usual prescription for closed superstring scattering
amplitudes in the pure spinor formalism. Since three-point amplitudes
of half-BPS states should be independent of $r_{AdS}$, the computation
of these three-point amplitudes should be similar in the topological string
prescription and in the pure spinor formalism. 

In a flat background using the pure spinor formalism, integration
over the left and right-moving worldsheet zero modes implies
that non-vanishing correlation functions require 3 $\l$'s and 3 $\lh$'s
as well as 5 $\t$'s and 5 $\th$'s in the combination \pureone
\eqn\zeroflat{(\l\g^m\t)(\l\g^n\t)(\l\g^p\t)(\t\g_{mnp}\t)
(\lh\g^q\th)(\lh\g^r\th)(\lh\g^s\th)(\th\g_{qrs}\th).}
In an $AdS_5\times S^5$ background using the pure spinor formalism,
it will be argued that the analogous zero mode measure factor is
simply 
\eqn\zeroads {(\eta_{\a\bh}\l^\a\lh^\bh)^3}
where $\eta_{\a\bh}\equiv \g^{01234}_{\a\bh}$. Moreover, for half-BPS
states, 
the unintegrated closed string vertex operator is 
\eqn\closedvertex{V = (\eta_{\a\bh}\l^\a \lh^\bh) f(x,\t,\th) + ...}
where $...$ is determined by BRST invariance. Since the three-point
tree amplitude prescription using the pure spinor formalism is
${\cal A} = \langle V_1 V_2 V_3 \rangle$,
one finds that after integrating over the pure spinor ghosts using
the measure factor of \zeroads, 
the pure spinor ghosts
trivially decouple and the pure spinor computation reduces to
the topological amplitude computation.

In section 2 of this paper, the topological A-model of \limone\limtwo\ 
is reviewed and is shown to be the gauge-fixed version of a 
${\cal G}/{\cal G}$ principal chiral model. In section 3, topological
amplitudes in this model are shown to compute super-Yang-Mills
Feynman diagrams in the `t Hooft large-N expansion. 
And in section 4, these topological amplitude computations are
compared with closed superstring amplitude computations using
the pure spinor formalism in an $AdS_5\times S^5$ background.

\newsec{Topological $AdS_5\times S^5$ Sigma Model}

\subsec{Review of ${{PSU(2,2|4)}\over{SU(2,2)\times SU(4)}}$
coset model}

In \limone\ and \limtwo, 
the pure spinor version of the superstring action in
an $AdS_5\times S^5$ background was mapped to an $N=(2,2)$
worldsheet supersymmetric sigma model based on the coset 
${{PSU(2,2|4)}\over{SU(2,2)\times SU(4)}}={{U(2,2|4)}\over{U(2,2)
\times U(4)}}$. Note that before introducing worldsheet gauge fields,
the non-linear sigma model based on the coset
${{PSU(2,2|4)}\over{SU(2,2)\times SU(4)}}$ is equivalent to the
non-linear sigma model based on the coset ${{U(2,2|4)}\over{U(2,2)
\times U(4)}}$. It was more convenient in \limtwo\ to use the coset
${{U(2,2|4)}\over{U(2,2)
\times U(4)}}$ since the $U(1)$ gauge field of $U(4)$ was necessary
for expressing the action as a gauged linear sigma model. In this paper,
the gauged linear sigma model will not play any role and it will be
necessary to use the coset
${{PSU(2,2|4)}\over{SU(2,2)\times SU(4)}}$ so that the worldsheet
gauge symmetries do not include the ``bonus'' $U(1)$ symmetry.

This non-linear sigma model was constructed from
a fermionic coset $G$ taking values in 
${{PSU(2,2|4)}\over{SU(2,2)\times SU(4)}}$
together with the bosonic ghosts $[Z^A_J, Y_A^J, \bar Z^J_A, \bar Y_J^A]$
where $A=1$ to 4 is an $SU(2,2)$ index and $J=1$ to 4 is an $SU(4)$ index.
The coset $G$ can be parameterized as $G(\t,\th)
= e^{\t^\a Q_\a + \th^\ah Q_\ah}$
where $\t^\a$ and $\th^\ah$ are 32 fermionic worldsheet scalars and,
after performing an $A$-twist, $(Z^A_J,\bar Z_A^J)$ are worldsheet
scalars and $Y_A^J$ and $\bar Y^A_J$ carry conformal weight $(1,0)$
and $(0,1)$. 

The map between these variables and the worldsheet variables of
the pure spinor formalism can be found in \limone\limtwo\
and will not be necessary
here. Up to a BRST-trivial term, this map takes the pure spinor
version of the $AdS_5\times S^5$ sigma model into the worldsheet action
\eqn\acttwo{S= r^2_{AdS} \int d^2 z [
(G^{-1}\p G)^J_A (G^{-1}\pb G)_J^A}
$$
-Y^J_A [\pb Z^A_J + 
(G^{-1}\pb G)^A_B Z^B_J
- (G^{-1}\pb G)^K_J Z^A_K ]
+\bar Y_J^A [\p \bar Z_A^J - 
(G^{-1}\pb G)^B_A \bar Z_B^J
+ (G^{-1}\pb G)_K^J \bar Z_A^K]$$
$$+
Y^J_A Z^A_K \bar Z^K_B \bar Y^B_J -
Z^A_J Y^J_B \bar Y^B_K \bar Z^K_A]. $$
Although one can combine $(\t^\a,\th^\ah,Z^A_J,\bar Z^J_A,Y_A^J,\bar Y^J_A)$
into $N=(2,2)$ worldsheet superfields and write \acttwo\ in worldsheet
superspace, it will be more convenient here to leave the 
worldsheet action in components.

It will be useful to note that by introducing the $SU(2,2)\times SU(4)$
worldsheet gauge fields $(A^A_B,\bar A^A_B)$ and $(A^J_K,\bar A^J_K)$,
\acttwo\ can be written as
\eqn\actthree{S= r^2_{AdS} \int d^2 z [
(G^{-1}\p G)^J_A (G^{-1}\pb G)_J^A + (G^{-1}\p G - A)^A_B (G^{-1}\pb G -\bar
A)^B_A
}
$$- (G^{-1}\p G - A)^J_K (G^{-1}\pb G -\bar A)^K_J
-Y^J_A (\bar\nabla Z)^A_J  
+\bar Y_J^A (\nabla\bar Z)_A^J ] $$
where 
$(\bar\nabla Z)^A_J = \pb Z^A_J + \bar A^A_B Z^B_J
- \bar A^K_J Z^A_K$ and 
$
(\nabla \bar Z)^J_A = \p \bar Z^J_A - A^B_A \bar Z_B^J
+ A_K^J \bar Z_A^K.$
Although not manifest when written in components, \actthree\ has
$N=(2,2)$ worldsheet supersymmetry and the $N=(2,2)$ worldsheet
superconformal generators are 
\eqn\superconf{Z^A_J (G^{-1}\p G)^J_A,\quad 
Y_A^J (G^{-1}\p G)_J^A,\quad
\bar Y^A_J (G^{-1}\pb G)^J_A,\quad 
\bar Z_A^J (G^{-1}\pb G)_J^A.}
So after performing an $A$-twist, the BRST operator in this topological
$A$-model is identified with
\eqn\BRSTtop{Q = 
\int dz Z^A_J (G^{-1}\p G)^J_A +\int d\bar z \bar Z_A^J
(G^{-1}\pb G)^A_J .}

As explained in \limtwo, the BRST operator of \BRSTtop\ for the topological
$A$-model is not mapped into
the BRST operator of the pure spinor formalism whose cohomology
defines the physical spectrum at large $r_{AdS}$.
Nevertheless, it was conjectured that at small $r_{AdS}$, the BRST operator
of \BRSTtop\ can be used to define the physical states. This conjecture
recently gained support from a paper showing that half-BPS super-Yang-Mills
Wilson
loops are described by D-branes in this topological A-model \bonelli.

\subsec{Principal chiral model}

In this subsection, it will be shown that the action of \actthree\
together with the BRST operator of \BRSTtop\ can be understood as
a gauge-fixed version of the ${\cal G}/{\cal G}$
principal chiral model where ${\cal G} = PSU(2,2|4)$. So the pure
spinor version of the $AdS_5\times S^5$ sigma model can be mapped
into a ${\cal G}/{\cal G}$
principal chiral model. It will be also be shown that other
gauge fixings of the 
${\cal G}/{\cal G}$ principal chiral model give rise to models
based on the coset ${{PSU(2,2|4)}\over{SU(2|2)\times SU(2|2)}}$
or ${{PSU(2,2|4)}\over{SU(1,1|2)\times SU(1,1|2)}}$. Like the
${{PSU(2,2|4)}\over{SU(2,2)\times SU(4)}}$ coset, these cosets
are symmetric spaces and their actions
are conformally invariant at the quantum level.
However, unlike the 
${{PSU(2,2|4)}\over{SU(2,2)\times SU(4)}}$ coset which contains
32 fermions and no bosons, these cosets contain 16 bosons and 16 fermions.

The worldsheet action
for the ${\cal G}/{\cal G}$ principal chiral model is defined as
\eqn\actfour{S = r_{AdS}^2 \int d^2 z (g^{-1}\p g - A)^R_S
(g^{-1}\pb g - \bar A)^S_R}
where $g$ takes values in $PSU(2,2|4)$, $R=(A,J)$ is a $PSU(2,2|4)$ index,
and $(A_R^S,\bar A_R^S)$ are worldsheet gauge fields taking values in
the $PSU(2,2|4)$ Lie algebra. Naively, this action is trivial since
one can shift $(A^R_S,\bar A^R_S)$ to eliminate $g$. However, as
will be seen in the following section, non-trivial solutions can
be obtained by 
introducing a kinetic term ${1\over{e^2}}\int d^2 z F^R_S F_R^S$
for the worldsheet gauge field and taking the infrared limit 
$e\to \infty$ at the end of the computation.

The worldsheet action of \actfour\ has a local $PSU(2,2|4)$ gauge
invariance under which $\d g=g\Omega$ and $\d A = d\Omega + [A,\Omega]$.
To relate \actfour\ to the action of \actthree, 
one should gauge-fix the $SU(2,2)\times SU(4)$ subgroup of this
invariance by choosing the gauge 
$g = G(\t,\th) = e^{\t^a Q_\a + \th^\ah Q_\ah}$. Furthermore, one
should gauge-fix the remaning 32 fermionic invariances by choosing the
gauge 
\eqn\gaugefixing{A^J_A =0,\quad \bar A_J^A =0}
for the fermionic worldsheet gauge fields.

The gauge choice $g=G(\t,\th)$ does not require Faddeev-Popov ghosts,
however, the gauge choice of \gaugefixing\ requires the
Faddeev-Popov ghosts $(Z^A_J,\bar Z^J_A)$ and antighosts
$(Y_A^J,\bar Y_J^A)$ with the worldsheet action
\eqn\ghostaction{S_{ghost} = \int d^2 z  [- Y_A^J (\bar\nabla Z)_J^A
+\bar Y_J^A (\nabla \bar Z)^J_A ] }
where $(\bar\nabla Z)_J^A$ and
$(\nabla\bar Z)^J_A$ are defined below \actthree.
So after gauge-fixing, the worldsheet action is 
\eqn\actfive{S= r^2_{AdS} \int d^2 z [
(G^{-1}\p G)^J_A (G^{-1}\pb G)_J^A + 
(G^{-1}\p G - A)^A_J (G^{-1}\pb G -\bar A)^J_A
}
$$+ (G^{-1}\p G - A)^A_B (G^{-1}\pb G -\bar A)^B_A 
- (G^{-1}\p G - A)^J_K (G^{-1}\pb G -\bar A)^K_J 
-Y^J_A (\bar\nabla Z)^A_J 
+\bar Y_J^A (\nabla\bar Z)_A^J ]. $$
Assuming that the kinetic term ${1\over{e^2}}\int d^2 z F^A_J F^J_A$
for the fermionic
gauge fields $A_J^A$ and $\bar A^J_A$ can be ignored in the limit
$e\to \infty$, one can
integrate out these fermionic gauge fields to obtain the action 
of \actthree. Furthermore, the standard BRST quantization
method implies that the
BRST operator arising from the
gauge-fixing of \gaugefixing\ is precisely \BRSTtop.

So the ${{PSU(2,2|4)}\over{SU(2,2)\times SU(4)}}$ worldsheet
action and BRST operator can be understood as coming from the
${\cal G}/{\cal G}$ principal chiral model in the gauge
$A^J_A = \bar A^A_J=0$. If one had instead chosen the gauge
\eqn\gaugetwo{A^a_{j'}=A^j_{\dot a} = A^a_{\dot a} = A^j_{j'}=0,\quad
\bar A_a^{j'}=\bar A_j^{\dot a} = \bar A_a^{\dot a} = \bar A_j^{j'}=0,}
where the $SU(2,2)$ and $SU(4)$ indices have been split into
$SU(2)\times SU(2)$ and $SU(2)\times SU(2)$
indices as $A=(a,\dot a)$ and $J=(j,j')$ for $a,\dot a,j,j'=1$ to 2,
the resulting action and BRST operator would be constructed in
a similar manner to \actthree\ using the coset ${{PSU(2,2|4)}\over
{PS[U(2|2)\times
U(2|2)]}}$. 
Similarly, if one had split the $SU(2,2)$ and $SU(4)$ indices into
$SU(1,1)\times SU(1,1)$ and $SU(2)\times SU(2)$ indices, the resulting
action and BRST operator would be constructed using the coset
${{PSU(2,2|4)}\over{PS[U(1,1|2)\times
U(1,1|2)]}}$.

So by starting with the ${\cal G}/{\cal G}$ principal chiral model
and choosing different gauge-fixings, one can relate topological A-models
based on different symmetric coset spaces. Since the denominator of
the coset determines the manifest symmetries, the worldsheet actions
based on the 
${{PSU(2,2|4)}\over{PS[U(2|2)\times
U(2|2)]}}$ and 
${{PSU(2,2|4)}\over{PS[U(1,1|2)\times U(1,1|2)]}}$ cosets may be useful
for describing BPS states which preserve different symmetries than
the half-BPS Wilson loops described in \bonelli.

\newsec{Feynman Diagrams from Topological Model}

\subsec{Physical observables}

As explained in the previous section, the topological A-model of 
\limone\limtwo\ 
can be understood as a gauge-fixed version of the ${\cal G}/{\cal G}$
principal chiral model whose worldsheet action is 
\eqn\actsix{S = 
Tr \int d^2 z [r_{AdS}^2 (g^{-1}\p g - A)
(g^{-1}\pb g - \bar A) + {1\over{e^2}} F^2 ]}
where $g$ takes values in $PSU(2,2|4)$, $(A,\bar A)$ is
a $PSU(2,2|4)$ worldsheet gauge field with field strength $F$,
and one takes the infrared limit $e\to \infty$ at the 
end of the computation. Naively, this model
has no physical states since one can use the local $PSU(2,2|4)$
symmetry to gauge $g=1$ and, in the limit $e\to \infty$, the gauge field
does not propagate.

Since the mass of the gauge field is $e ~r_{AdS} $, the fluctuations of
the gauge field have size of order $(e~ r_{AdS})^{-1}$.
If $r_{AdS}$ is not small, the size of the fluctuations goes quickly
to zero in the
infrared limit $e\to\infty$. However, if $r_{AdS}$ is infinitesimal, these
fluctuations may not be small and one can consider ``holes'' of
size $(e~ r_{AdS})^{-1}$ in the worldsheet where the gauge field is nonzero.

Physical observables will be related to these fluctuations of the gauge
field, and the locations of the ``holes'' will correspond to the locations
of closed string vertex operators which carry global $PSU(2,2|4)$ indices. 
Since physical observables must
be gauge invariant with respect to the local $PSU(2,2|4)$ symmetry,
one needs to construct gauge-invariant operators out of $g$ and $A$ which
describe these physical observables. 

Under local $PSU(2,2|4)$ transformations parameterized by $\Omega^{I'}_{J'}$, 
the coset $g_I^{I'}$
and the gauge field $A^{I'}_{J'}$ transform as
\eqn\transform{\d g_I^{I'} = g_I^{J'}\Omega_{J'}^{I'},
\quad
\d A^{I'}_{J'} = d\Omega^{I'}_{J'} + A^{K'}_{J'} \Omega^{I'}_{K'} -
\Omega^{K'}_{J'} A^{I'}_{K'},}
where $I$ is a global $PSU(2,2|4)$ index and $I'$ is a local
$PSU(2,2|4)$ index. And under global $PSU(2,2|4)$ transformations parameterized
by $\Sigma^I_J$,
\eqn\globalg{\d g_I^{I'} = \Sigma_I^J g_J^{I'}, \quad \d A^{I'}_{J'}=0.}
In general, the indices $I$ and $I'$ could label any representation of
$PSU(2,2|4)$, however, throughout the rest of 
this paper the indices $I$ and $I'$
will always denote the ``singleton'' representation corresponding
to the on-shell states of a super-Maxwell multiplet. 
The singleton representation is infinite-dimensional and it will be convenient
to use the label $I=Z$ to denote the onshell scalar at zero momentum
with $+1$ R-charge in the $56$ direction of $SO(6)$. All other states
in the singleton representation can be obtained by repeatedly
applying $PSU(2,2|4)$
transformations on this $I=Z$ state.

$PSU(2,2|4)$ gauge-invariant operators will be constructed with the help
of the
$PSU(2,2|4)$-invariant tensors $\d_{IJ}$ and $\e_{IJK}$ where $I,J,K$
indices always denote the singleton representation.
If the index $I$ denotes the super-Yang-Mills state $\phi_I$, the tensors
$\d_{IJ}$ and $\e_{IJK}$ are defined to be the free propagator and the bare
three-vertex of super-Yang-Mills as
\eqn\deftensor{\d_{IJ} = \langle \phi_I \phi_J \rangle, \quad
\e_{IJK} = \langle \phi_I \phi_J \phi_K \rangle}
where the color indices of $\phi_I$ are ignored.
An explicit construction of $\d_{IJ}$ 
can be found in section (6.2) of \ref\mikhone
{A. Mikhailov, {\it Notes on Higher Spin Symmetries},
hep-th/0201019.} and section (3.1) of \ref\narain{L. Alday,
J. David, E. Gava and K.S. Narain, {\it Towards a String Bit 
Formulation of $N=4$ Super-Yang-Mills}, JHEP 0604 (2006) 014,
hep-th/0510264.} where states in the singleton representation
are mapped using a non-unitary transformation
into states in position space. Once the
singleton states are described in position space, one can use the standard
definitions of the propagator $\d_{IJ}$ and three-vertex $\e_{IJK}$.
It will also be useful
to define the tensor $\d^{IJ}$ to be the inverse of $\d_{IJ}$
which corresponds to the super-Yang-Mills kinetic operator.
For example, if the indices $I$ and $J$ correspond to
the scalars $Z_{[ij]}(x)$ and $Z_{[kl]}(y)$ where $i,j,k,l=1$ to 4 are $SU(4)$
indices and $x^m$ and $y^m$ label the point in $d=4$, 
\eqn\defpr{\d_{IJ} = \e_{ijkl}(x-y)^{-2}\quad
{\rm and}\quad \d^{IJ} = \e^{ijkl}\p_m\p^m\delta^4(x-y).}
And if $I$ and $J$ correspond to the chiral gluinos $\psi_i^\a(x)$ and
$\psi_j^\b(y)$ and $K$ corresponds to the scalar $Z_{[kl]}(z)$, 
\eqn\defepr{\e_{IJK} = 
\e_{ijkl} \e_{\dot\a\bd} \int d^4 w~ F^{\a\dot\a}(x-w) 
~~F^{\b\bd}(y-w)~~ G(z-w)}
where $F^{\a\dot\a}(x-w) = \s_m^{\a\dot\a} (x-w)^m (x-w)^{-4}$ is the spinor
propagator and $G(z-w)=(z-w)^{-2}$ is the scalar propagator.

Note that
when expressed in terms of on-shell plane-wave states, these
$PSU(2,2|4)$-invariant tensors either vanish or become singular.
For example, 
$\d_{IJ} = p^{-2} \d^4(p+q)$ and $\d^{IJ}= p^2 \d^4(p+q)$
when expressed in terms of plane-wave scalar
states with momenta $p_m$ and $q_m$.
To resolve these singularities, one needs to introduce a
regulator which plays the role of the usual $(i\e)$ prescription
in Feynman rules. Furthermore, one needs to convert sums over
singleton indices into integrals over internal
off-shell momenta. At the moment, it is unclear how to do this in
a natural way.\foot{I would like to thank Andrei Mikhailov and Warren
Siegel for discussions on this point.}

Up to overall normalization factors, $\d_{IJ}$ and $\e_{IJK}$ are
the only independent $PSU(2,2|4)$ invariant tensors that can be constructed
from the singleton representation. This follows from the fact that the
$N=4$ $d=4$ super-Yang-Mills action is the unique $PSU(2,2|4)$-invariant
action, and the overall normalization of $\d_{IJ}$ and $\e_{IJK}$
can be absorbed by rescaling the super-Yang-Mills fields and the 
super-Yang-Mills coupling constant. Note that $\d_{IJ}$ is invariant
under the ``bonus'' $U(1)$ symmetry which enlarges $PSU(2,2|4)$
to $U(2,2|4)$, however
$\e_{IJK}$ is not invariant under the ``bonus'' $U(1)$ and is invariant
only under $PSU(2,2|4)$.

At each ``hole'' in the worldsheet, 
the fluctuations of size $(e~ r_{AdS})^{-1}$
will be represented by a closed string vertex operator
which carries global $PSU(2,2|4)$ indices and corresponds to a gauge-invariant
super-Yang-Mills operator. At zero coupling constant, the gauge-invariant
super-Yang-Mills operator can be described as a spin chain of $L$
singleton representations which is invariant under cyclic permutations.
Note that at zero coupling constant, $PSU(2,2|4)$ transformations act
linearly on the super-Yang-Mills fields so that each singleton representation
describes a single super-Yang-Mills field.

The closed string vertex operator at the $r^{th}$ hole will have the form
\eqn\closedv{V_{r}(z_r) = f^{I_1 ... I_{L_r}}_{r} V_{I_1 ... I_{L_r}}(z_r)}
where $V_{I_1 ... I_{L_r}}(z_r)$ is the vertex operator for the spin chain
with $L_r$ singleton representations and $f_{r}^{I_1 ... I_{L_r}}$ 
are the ``polarizations'' of the fields in the $r^{th}$ spin chain.
Since $V_{I_1 ... I_{L_r}}(z_r)$ carries $L_r$ global $PSU(2,2|4)$
indices and is constructed from $g_I^{I'}$ and $A^{I'}_{J'}$, the only
possibility is that 
$V_{I_1 ... I_{L_r}}(z_r)$ is proportional to 
$g_{I_1}^{{I'}_1}(z_r) ...
g_{I_{L_r}}^{{I'}_{L_r}}(z_r)$.

In order to construct a physical observable which is invariant under
local $PSU(2,2|4)$ transformations, each of the $L_r$ primed indices
${I'}_1 ... {I'}_{L_r}$ must be contracted with a path-ordered Wilson-line
operator $P(\exp \int_{z_r} A)_{I'}^{J'}$ where the endpoint of 
the Wilson-line operator will be determined shortly. Furthermore,
the $L_r$ Wilson lines emerging from $z_r$ will be prohibited from crossing
and will be ordered clockwise such that they preserve the order of the indices
on $V_{I_1 ... I_{L_r}}$. This clockwise ordering implies that the vertex 
operator
\eqn\cyclical{V_{I_1 ... I_{L_r}} = g_{I_1}^{{I'}_1}(z_r)  
(P e^{\int_{z_r} A})_{{I'}_1}^{{J'}_1} ~~ 
g_{I_2}^{{I'}_2}(z_r) 
(P e^{\int_{z_r} A})_{{I'}_2}^{{J'}_2} ~~ 
... ~~ 
g_{I_{L_r}}^{{I'}_{L_r}}(z_r)  
(P e^{\int_{z_r} A})_{{I'}_{L_r}}^{{J'}_{L_r}}  }
is invariant under cyclic permutations of the indices $I_1 ... I_{L_r}$.
The requirement that Wilson lines do not cross will be treated
as an assumption, but the assumption might be justified by the presence
of singularites of crossing Wilson lines 
before taking the infrared limit $e\to\infty$.

Finally, to construct a gauge-invariant observable, one needs to contract
the remaining $J'$ index on each of the $L_r$ Wilson lines which emerge
from the $r^{th}$ hole. These $J'$ indices will be contracted either by
joining the endpoints of two Wilson lines and contracting their $J'$ and
$K'$ indices with the $PSU(2,2|4)$-invariant
tensor $\d_{J'K'}$, or by joining the endpoints
of three Wilson lines and contracting their $J'$, $K'$ and $L'$ indices
using the $PSU(2,2|4)$-invariant
tensor $\e_{J'K'L'}$. In the first case, the Wilson lines
resemble a Feynman propagator connecting two super-Yang-Mills fields
and, in the second case, the Wilson lines resemble a cubic vertex
connecting three super-Yang-Mills fields. One can also construct
gauge-invariant observables involving ``internal'' Wilson lines
where both endpoints of the Wilson line are contracted with 
$PSU(2,2|4)$-invariant tensors.\foot{Andrei Mikhailov has pointed
out that this network of Wilson lines resembles the
network of transfer matrices considered in \ref\mikhtwo
{A. Mikhailov and S. Schafer-Nameki, {\it Algebra of
Transfer-Matrices and Yang-Baxter Equations on the String Worldsheet
in $AdS_5\times S^5$}, arXiv:0712.4278[hep-th].}. It would be very interesting
to explore this relation, perhaps using the transfer matrices recently
constructed in \ref\vali{W. Linch, III and B.C. Vallilo, 
{\it Integrability of the Gauged Linear Sigma Model for $AdS_5\times S^5$},
arXiv:0804.4507[hep-th].}.}

\subsec{Feynman diagrams}

It will now be claimed that after taking the infrared limit
$e\to\infty$, this network of vertex operators connected
by Wilson lines reproduces the standard Feynman diagram computation
in the 't Hooft large $N$ expansion of perturbative super-Yang-Mills.
Since the Wilson lines are prohibited from crossing on the worldsheet,
the network of Wilson lines on a worldsheet of genus $g$ corresponds
to a thickened Feynman diagram of genus $g$. In the `t Hooft large $N$ limit,
the thickened Feynman diagram of genus $g$ with $F$ faces
contributes a factor proportional to 
\eqn\thooft{
N^{2-2g} (\l_{'t Hooft})^{F+2g-2} =
(\l^2_{YM})^{2g-2} (\l^2_{YM} N)^F}
where $\l_{'t Hooft} = \l^2_{YM} N$. 
Since $\l_{string} = \l^2_{YM}$ and the genus $g$ closed string amplitude
is proportional to $(\l_{string})^{2g-2}$, the factor of 
\thooft\ is reproduced if each face contributes a factor of $\l_{'t Hooft}$.
Note that unlike the Chern-Simons/conifold duality where faces correspond
to holes on the worldsheet, faces in this network are the regions
bounded by Wilson lines and do not
correspond to holes on the worldsheet.\foot{I would like to thank
Rajesh Gopakumar for stressing this point.}

Extending the Maldacena conjecture to small $r_{AdS}$ would imply that
each face should contribute a factor of $\l_{'t Hooft} = r_{AdS}^4$.
Although not rigorous, an argument which implies precisely such a
contribution is as follows: After using the local $PSU(2,2|4)$ symmetry
to gauge-fix $g_I^{I'} = \d_I^{I'}$, the worldsheet action in the limit
$e\to\infty$ is simply
\eqn\limitact{S = r^2_{AdS}\int d^2 z A^{I'}_{J'} \bar A_{I'}^{J'}.}
If one assumes that $A^{I'}_{J'}$ can be discontinuous when crossing
a Wilson line, the number of zero modes of $A^{I'}_{J'}$ is equal to
the number of faces in the network. Furthermore, the action of
\limitact\ implies that integration over each
bosonic zero mode of $A$ produces a factor of $(r^2_{AdS})^{-1}$ and
integration over each fermionic zero mode of $A$ produces a factor
of $(r^2_{AdS})^{+1}$. Since the $PSU(2,2|4)$ Lie algebra has 30
bosonic generators and 32 fermionic generators, the net contribution
is a factor of $r_{AdS}^4$ for each face in the network. Note that for this
argument to work, it is crucial that the gauge group is chosen
to be $PSU(2,2|4)$ as opposed to
$U(2,2|4)$, and this choice is also required by the fact that $\e_{IJK}$
is not invariant under the bonus $U(1)$ symmetry.

Up to some subtleties mentioned at the end of this section, one can
also argue that the network of Wilson lines connecting the vertex operators
$V_{I_1 ... I_{L_r}}(z_r)$ contributes to the topological amplitude using the
same rules as the Feynman diagram connecting the gauge-invariant 
super-Yang-Mills operators described by $V_{I_1 ... I_{L_r}}$.
In the limit where $e\to \infty$, the equation of motion for the gauge field
is $A = g^{-1} dg$. So after taking the limit
$e\to\infty$, the path-ordered Wilson line
operator connecting $g(y)$ and $g^{-1}(z)$ contributes
\eqn\pathon{g_I^{I'}(y) ~P(e^{\int_y^z A})_{I'}^{J'} ~(g^{-1}(z))_{J'}^J =
g_I^{I'}(y)~ P(e^{\int_y^z g^{-1}dg})_{I'}^{J'}~ (g^{-1}(z))_{J'}^J = \d_I^J.}
So the network of Wilson lines which connect the $M$ vertex operators
$V_r(z_r)$
$ = f_r^{I_1 ... I_{L_r}} V_{I_1 ... I_{L_r}}(z_r)$ contributes
the topological amplitude
\eqn\contamp{{\cal A} = \l_{string}^{2g-2}~ (r_{AdS}^4)^{faces} ~~
(\prod_{r=1}^M 
f_r^{I^{(r)}_1 ... I^{(r)}_{L_r}})~~ T_{I^{(1)}_1 ... I^{(1)}_{L_1}
~
I^{(2)}_2 ... I^{(2)}_{L_2}
~
 ...
~
I^{(M)}_1 ... I^{(M)}_{L_M}} }
where $T_{I^{(1)}_1 ... I^{(M)}_{L_M}}$ is a $PSU(2,2|4)$ invariant
tensor containing $\sum_{r=1}^M L_r$ indices which is constructed
from the $PSU(2,2|4)$-invariant tensors $\d_{IJ}$, $\e_{IJK}$ and $\d^{IJ}$.
Since $\d_{IJ}$ and $\e_{IJK}$ correspond to the propagator and
three-vertex of super-Yang-Mills, the tensor $T$ computes the contribution
of the super-Yang-Mills Feynman diagram which is described by the 
Wilson-line network. 
As expected from a topological amplitude computation, the amplitude of
\contamp\ is independent of the locations of the vertex operators and
only depends on the topology of the Wilson-line network.

Using the above arguments, it seems reasonable to conjecture that
the topological amplitude for
the network of Wilson lines correctly
reproduces the perturbative computation of gauge-invariant
super-Yang-Mills correlation functions.
However, there are several possible subtleties in proving this conjecture
which deserve further study.
Firstly, covariant Feynman diagram computations require gauge-fixing
and ghosts, and the tensor $T$ of \contamp\ should somehow automatically
include the ghost contributions. Secondly, loop computations require
regularization, and one expects that a similar regularization for the
tensor $T$ is necessary when one has multiply contracted indices such
as $\e_{IJK} \d^{KL} \e_{LMN} \d^{NI}$.
Thirdly, the quartic vertex of super-Yang-Mills Feynman diagrams should
somehow arise in $T$ from a contact term when evaluating the contribution
$\e_{IJK}\d^{KL}\e_{LMN}$ that arises from the contraction of two 
cubic vertices. Note that after introducing auxiliary fields,
the super-Yang-Mills action can be written as a cubic action.
So it would not be surprising if the quartic vertex could be interpreted
as a contact term of two cubic vertices coming from integrating
out the auxiliary field.

\newsec{Comparison with Superstring Amplitudes}

\subsec{$AdS_5\times S^5$ measure factor}

In the previous section, it was argued that perturbative super-Yang-Mills
correlation functions can be computed as topological amplitudes using
the small radius limit of the topological $AdS_5\times S^5$ sigma model.
These topological amplitude computations naively look very different
from closed superstring amplitude computations using the pure spinor
formalism. For example, in a flat background, unintegrated closed
superstring vertex operators for supergravity states have the form
$V = \l^\a \lh^\ah A_{\a\ah}(x,\t,\th)$ where $\l^\a$ and $\lh^\ah$
are the left and right-moving pure spinor ghosts. And three-point
amplitudes in a flat background are computed by
${\cal A} = \langle V_1 V_2 V_3 \rangle$ using the zero mode
measure factor
\eqn\zeromeasure{\langle (\l\g^m\t)(\l\g^n\t)(\l\g^p\t)(\t\g_{mnp}\t)
(\lh\g^q\th)(\lh\g^r\th)(\lh\g^s\th)(\th\g_{qrs}\th)\rangle =1.}

Since supergravity states in an $AdS_5\times S^5$ background correspond
to half-BPS super-Yang-Mills gauge-invariant operators, one expects
that the three-point amplitude for these states should be independent
of $r_{AdS}$. So it should be possible to relate the topological amplitude
of this three-point half-BPS correlation function at small radius with
the superstring amplitude computation at large radius. In this section,
it will be shown how to relate these two computations.

The first step in relating the two computations is to determine
the zero mode measure factor using the pure spinor
formalism for the superstring
in an $AdS_5\times S^5$ background.
This measure factor should be in the BRST cohomology at ghost-number
$(3,3)$ where the left and right-moving BRST operators are \adsqt
\eqn\brst{Q = \int dz~\eta_{\a\bh}\l^\a (g^{-1}\p g)^\bh,\quad
\bar Q = \int d\bar z~\eta_{\a\bh}\lh^\bh (g^{-1}\pb g)^\a,}
$\eta_{\a\bh}=\g^{01234}_{\a\bh}$, and $g$ takes values in the
${{PSU(2,2|4)}\over{SO(4,1)\times SO(5)}}$ coset. Under the BRST
transformations generated by \brst, 
\eqn\brsttransf{\d g = g (\l^\a T_\a + \lh^\ah T_\ah), \quad \d \l^\a =0,\quad
\d \lh^\ah =0,}
where $T_\a$ and $T_\ah$ are the 32 fermionic generators of $PSU(2,2|4)$.

One clue in constructing the zero mode measure factor in an $AdS_5\times S^5$
background is to note that for the Type IIA superstring in a flat
background, the measure factor of 
\zeromeasure\ can be written as 
\eqn\zerotwo{ \langle ~[(\l\g^m\t)(\lh\g_m\th)]^5 (\l^\a\lh_\a)^{-2}~
\rangle =1}
using the identities
\eqn\iden{ (\l\g^{m_1}\t) (\l\g^{m_2}\t) (\l\g^{m_3}\t) (\l\g^{m_4}\t)
(\l\g^{m_5}\t) = (\l\g^{m_1 ... m_5}\l)
(\l\g^n\t)(\l\g^p\t)(\l\g^q\t)(\t\g_{npq}\t)}
and $(\l\g^{m_1 ... m_5}\l)(\lh\g_{m_1 ... m_5}\lh) = (\l^\a\lh_\a)^2$
where overall proportionality factors are being ignored.

The operator $V_{flat} = (\l\g^m\t)(\lh\g_m\th)$ appearing in \zerotwo\ is
the vertex operator of the graviton trace at zero momentum, and is
related to the worldsheet Lagrangian $L_{flat}$ in a flat background 
by 
\eqn\relatflat{Q \bar Q L_{flat} = \p \pb V_{flat}.} 
Using the worldsheet Lagrangian $L_{AdS}$ for the pure spinor formalism
in an $AdS_5\times S^5$ background, 
one can similarly
compute the vertex operator $V_{AdS}$
for the
$AdS$ radius modulus at zero momentum and one finds that 
\eqn\relatads{Q \bar Q L_{AdS} = \p \pb V_{AdS}} 
where $V_{AdS} = \eta_{\a\bh} \l^\a \lh^\bh$.

By analogy with the zero mode measure factor of \zerotwo, the natural
guess for the zero mode measure factor in an $AdS_5\times S^5$
background is therefore 
\eqn\zerotwo{ 
\langle ~(\eta_{\a\bh}\l^\a\lh^\bh)^5 (\eta_{\g\dh}\l^\g\lh^\dh)^{-2}~\rangle=
\langle ~(\eta_{\a\bh}\l^\a\lh^\bh)^3 ~\rangle =1. }
So unlike in a flat background, the $AdS_5\times S^5$
measure factor only involves the pure
spinor ghosts and does not involve the matter fields.
To verify that \zerotwo\ 
is the correct
measure factor, one can easily compute the tree amplitude of three
radius moduli described by the vertex operator 
$V_{AdS}=\eta_{\a\bh}\l^\a\lh^\bh$ and one finds that 
\eqn\ampv{{\cal A} =
\langle V_{AdS} V_{AdS} V_{AdS}\rangle =1.} 
Note that in a flat background, the analogous amplitude involving
the zero momentum graviton trace vanishes since $(V_{flat})^3$
contains 3 $\t$'s and 3 $\th$'s whereas the measure factor of 
\zeromeasure\ requires 5 $\t$'s and 5 $\th$'s. This result is
consistent with the fact that the $d=10$ effective action vanishes
in a flat background. But in an $AdS_5\times S^5$ background,
the effective action is a non-vanishing function of the $AdS$ radius.

\subsec{$AdS_5\times S^5$ vertex operators}

The next step in relating the computations of three-point
half-BPS amplitudes is to contruct the vertex operator for
a general supergravity state in the pure spinor formalism. 
As explained in \adsquant, one method for constructing the supergravity
vertex operators uses a bispinor superfield $A_{\a\bh}(x,\t,\th)$
satisfying the on-shell conditions
\eqn\onshell{ \g_{mnpqr}^{\a\g} \nabla_\g A_{\a\bh} =
\g_{mnpqr}^{\bh\gh} \nabla_\gh A_{\a\bh} = 0}
where $\nabla_\a$ and $\nabla_\ah$ are the covariant fermionic derivatives
in an $AdS_5\times S^5$ background. As in a flat background, the
unintegrated supergravity vertex operator in an $AdS_5\times S^5$
background can be expressed in terms of $A_{\a\bh}$ as
$V=\l^\a \lh^\bh A_{\a\bh}(x,\t,\th)$ and the on-shell conditions
of \onshell\ imply that $QV=\bar Q V =0$.

{}From the analysis in the previous subsection, it is clear that
the $\t=\th=0$ component of $\eta^{\a\bh} A_{\a\bh}$ is the radius
modulus, and other fields in the supergravity
multiplet can be obtained from this
modulus by supersymmetry transformations.
For example, the vertex operator for the scalar with $J$ units
of R-charge 
in the $56$ direction is
\eqn\vnew{V_J = (\eta_{\a\bh}\l^\a\l^\bh) a^{\pm J} e^{i J y_{56}} + ...}
where $a$ is the $x^5$ direction in $AdS_5$, $y_{56}$ is the $56$ direction
in $S^5$, the choice of $\pm$ sign determines the $AdS_5$ boundary condition
of the state, and $...$ contains terms higher order in $(\t,\th)$
which are determined by BRST invariance.

If the plus sign is chosen in \vnew\ so that $V_J$ diverges as $a\to\infty$,
the supergravity vertex operator corresponds to the $PSU(2,2|4)$
representation with $|J|$ lowered indices. Using the notation where
$I=Z$ corresponds to the zero-momentum scalar with $+1$ R-charge in the
56 direction and
$I=\bar Z$ corresponds to the zero-momentum scalar with $-1$ R-charge
in the 56 direction,
$V_J = V_{Z ... Z}$ when $J$ is positive and $V_J = V_{\bar Z ... \bar Z}$
when $J$ is negative. On the other hand, if the minus sign is chosen
in \vnew\ 
so that $V_J$ goes to zero as $a\to \infty$, the supergravity state
corresponds to the $PSU(2,2|4)$ representation with $|J|$ raised
indices. Defining $\d^{IJ}$ to be the same $PSU(2,2|4)$-invariant
tensor defined earlier, $V_J = V^{\bar Z ... \bar Z} = \d^{\bar Z I_1} ...
\d^{\bar Z I_J} V_{I_1 ... I_J}$ when $J$ is positive and
$V_J = V^{Z ... Z} = \d^{Z I_1} ...
\d^{Z I_{|J|}} V_{I_1 ... I_{|J|}}$ when $J$ is negative.

\subsec{Three-point supergravity amplitude}

Using the superstring vertex operators $V_J$ of \vnew,
it is easy to
compare the three-point superstring tree amplitudes of these states
with the topological
amplitude computations. For the amplitude 
\eqn\threea{{\cal A} = \langle V_{J_1}(z_1) V_{J_2}(z_2) V_{J_3}(z_3)\rangle,}
the measure factor of \zerotwo\ 
implies that ${\cal A}=1$ if and only if $J_1+ J_2 + J_3=0$ and if
the state with maximum $|J|$ charge has the opposite $AdS_5$ boundary
condition from the other two states. These conditions guarantee that
there are either an equal number of $Z$ subscript and $Z$ superscript
indices on the vertex operators, or an equal number of $\bar Z$
subscript and $\bar Z$ superscript indices. 

For example, suppose that $J_1$ is positive and $J_2$ and $J_3$ 
are negative such that $J_1+J_2+J_3=0$. If $V_{J_1}$ diverges when
$a\to\infty$, the amplitude 
$\langle V_{J_1}(z_1)V_{J_2}(z_2)V_{J_3}(z_3)\rangle=1$ implies that
\eqn\comput{
\langle ~V_{Z ... Z}(z_1)~~\d^{Z I_1} ...\d^{Z I_{|J_2|}}
V_{I_1 ... I_{|J_2|}} ~~
\d^{Z K_1} ...\d^{Z K_{|J_3|}}
V_{K_1 ... K_{|J_3|}}~ \rangle =1.}
To show that this result agrees with the topological amplitude
computation, note that for three-point amplitudes involving half-BPS
states, only the propagator contributes to the Feynman diagram
computation since the amplitude is independent of the super-Yang-Mills
coupling constant. Since there are no contributions from cubic
vertices, the
topological amplitude computation involves a single Wilson-line network
with $J_1$ propagators which contributes 
\eqn\topoamp{\langle ~V_{Z ... Z}(z_1)~ V_{I_1 ... I_{|J_2|}}(z_2) 
~V_{K_1 ... K_{|J_3|}}(z_3)~\rangle = \d_{Z I_1} ~ ...~ \d_{Z I_{|J_2|}}
~~~\d_{Z K_1} ~...~ \d_{Z K_{|J_3|}}.}
So using $\d_{ZI}\d^{IZ} =1$, one finds that \topoamp\ agrees with
\comput.

In comparing these topological amplitudes and pure spinor superstring
amplitudes, it was important that the $\l^\a$ and $\lh^\ah$ pure
spinor ghosts decoupled in a trivial manner in the superstring computation.
For amplitudes involving non-BPS states or more than three half-BPS
states, the pure spinor ghosts probably play a more complicated role
and it will be highly non-trivial to compare the two amplitude computations.
This is not surprising since these amplitudes are expected to have
non-trivial dependence on the $AdS$ radius. 

One situation which would be very interesting to study is the plane-wave
limit in which the external vertex operators carry large R-charge.
In this case, it might be possible to compare the topological and
superstring computations for a more general class of scattering
amplitudes. Perhaps in the limit of large R-charge,
the discrete set of contributions to the topological amplitude
combines into a continuous integral over worldsheet moduli in the
superstring amplitude computation.
Another speculation is that in the plane-wave limit, 8 bosonic and
8 fermionic components of the $PSU(2,2|4)$ gauge field might become
dynamical and reproduce the light-cone degrees of freedom of the superstring.

\vskip 1cm
{\bf Acknowledgment}

I would like to thank R. Gopakumar, P. Howe, V. Kazakov, R. Lopes de S\'a,
A. Mikhailov, N. Nekrasov, A. Polyakov,
W. Siegel, K. Skenderis, M. Staudacher, C. Vafa, B.C. Vallilo,
E. Witten, and especially J. Maldacena
for valuable discussions.
I would also like to thank the hospitality of the IAS at Princeton
where part of this research was done.
The research of N.B. was supported in part by
CNPq grant 300256/94-9 and FAPESP grant 04/11426-0.

\listrefs

\end